# Synergy Test for Antibacterial Activity: Towards the Research for a Consensus between the Fractional Inhibitory Concentration (Checkboard Method) and the Increase in Fold Area (Disc Diffusion Method)

**Mbarga Manga Joseph Arsene***

Department of Microbiology and Virology, Medical Institute, People Friendship University of Russia, Moscow, Russia

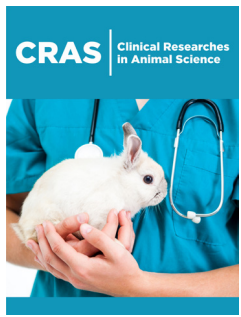

**\*Corresponding author:** Mbarga Manga Joseph Arsene, Department of Microbiology and Virology, Medical Institute, People Friendship University of Russia, Moscow, Russia

## Opinion

Antibiotic resistance is a topical problem for both humans and animals and has been the subject of special monitoring for two decades. Several recent studies, including ours, have shown that this phenomenon is accentuated by the transfer of certain genetic elements and cross resistance acquisition, the latter or both resulting from the misuse of these antimicrobial drugs [1-3]. A more careful use of antimicrobials, the search for new antibacterial compounds (including probiotics and phages) are the most recommended alternatives to overcome this situation [4,5]. However, tests for modulations of antimicrobial activity can also play a major role. The main goal of synergy studies is to assess whether substances with antibacterial properties can improve the effectiveness of existing antimicrobials or give them a second life against resistant germs. Moreover, recent studies have demonstrated the ability of silver nanoparticles [5] and extracts of certain plants [6] to boost the effectiveness of certain antibiotics. such as ampicillin, benzylpenicillin, cefazolin, ciprofloxacin, nitrofurantoin, and kanamycin. Yet, from these studies we found that there was a serious problem with the interpretation of the results when using the disk method with determination of the increase in fold area. Indeed, the use of this method firstly requires the determination of the diameter of inhibition of the antibiotic alone; follow by the determination of the combination of antibiotic + modulating substance (MS) or extract. Finally, the calculation of the increase in fold area by the formula [7]:

$$A = \frac{Y^2 - X^2}{X^2}$$

Where, "A" is the increase in fold area, "Y" the zone of inhibition for extract + antibiotic and "X" is the inhibition zone of antibiotic alone.

This disc method has the following drawbacks:

1) It does not take into account the diameter of inhibition of the MS.

2) Variations in the initial inhibition diameter mislead the results. Because if we refer to this formula, the small variations of the inhibition diameters will be interpreted (using the increase in fold area) as being very important for the initial inhibition diameters small compared to the same variations for larger inhibition diameters. big.

3) It is difficult to come out with the significance scales of the effect despite the calculation of the increase in fold area and even if it were done, this classification would still have to be dependent on the diameter ranges.

To circumvent these drawbacks, some authors who have used this method simply assumed that synergy existed when there was an increase of more than 4mm [8]. in the diameter of inhibition while others preferred to represent the results as histogram or tables using raw data. However, unlike modulation by the disc diffusion method, the checkboard method considered both minimum inhibitory concentrations of the two substances being tested. In this method, the following formula is used: FIC=FICA+FICB, with: $FICA = \frac{MIC'A}{MICA}$ and $FICB = \frac{MIC'B}{MICB}$; where FIC is the fractional inhibitory concentration, FIC A and FIC B are the FICs of each compound, MIC A and MIC B represent de MIC before the combination and MIC'A and MIC'B are the MIC of the same compounds after the combination [9,10].

The FIC index is interpreted as follows: FIC ≤0.5, synergy; 0.5 ≤FIC ≤1, addition of effects; 1≤ FIC≤4, indifference and for FIC> 4, Antagonism. In trying to highlight these two methods, for modulation by diffusion on disc, it would therefore be adequate for a better appreciation of the results, to consider a fourth variable representing the diameter of initial inhibition of the modulating substance. Thus, just as with the FIC, we could thus define the A1 and the A2 (respectively representing the increase in fold area of each of the two substances tested) in order to $A1 = \frac{Y-X}{X}$ and, $A1 = \frac{Y-Z}{Z}$, then finally calculate A=A1+A2. Consequently, several scenarios emerge for the interpretation of the results:

A. If A1 or A2 < 0 and A1 + A2> 0, there is antagonism, and this antagonism is caused by one of the two substances.

B. If A1 and A2<0 the two substances are categorically antagonistic.

C. If A1 and A2>0, there may be additional effect or synergy.

This last case is the most interesting and necessarily requires laboratory tests to assess from what positive value of "A" there would be a significant synergistic effect in correlation with other existing methods. Hence, the proposed approach has the advantage of considering the two substances used and even leaves place for the combination of more than 2 other substances. Additionally, this new approach of interpretation makes it possible to reduce the deviations of the values of "A" when facing same variations between the high inhibition diameters compared to the low ones.

Ultimately, in-depth studies are required to confirm or refute the hypothesis we describe here. In addition, though this hypothesis better describes the synergistic effect between 2 tested substances, further investigations would also be needed to establish the intervals and degrees of significance when there is synergy between the two tested substances.